\documentclass[twocolumn,superscriptaddress,floatfix,preprintnumbers,amssymb ,amsmath]{revtex4}
\usepackage{graphicx}
\usepackage{dcolumn}
\usepackage{bm}
\usepackage[latin1]{inputenc}
\usepackage[mathscr]{eucal}
\usepackage{epsfig}

\begin{document}
\title{Information-to-energy conversion in an electron pump}
\author{Dmitri V.\ Averin}
\affiliation{Department of Physics and Astronomy, Stony Brook University, SUNY, Stony Brook, NY 11794-3800, USA}
\author{Mikko M\"ott\"onen}
\affiliation{Low Temperature Laboratory, Aalto University, P.O. Box 13500, 00076 Aalto, Finland}
\affiliation{Department of Applied Physics/COMP, AALTO University, P.O. Box 14100, FI-00076 AALTO, Finland}
\author{Jukka P.\ Pekola}
\affiliation{Low Temperature Laboratory, Aalto University, P.O. Box 13500, 00076 Aalto, Finland}

\begin{abstract}
We propose and analyze a possible implementation of Maxwell's demon based on a single-electron pump. We show that measurements of the charge states of the pump and feedback control of the gate voltages lead to a net flow of electrons against the bias voltage ideally with no work done on the system by the gate control. The information obtained in the measurements converts thermal fluctuations into free energy. We derive the conditions on the detector back-action and measurement time necessary for implementing this conversion.
\end{abstract}

\date{\today}

\maketitle

Work, heat, and reversibility are topics of intense current interest in driven small systems, where fluctuations play an important role. Instead of familiar inequalities of macroscopic thermodynamic quantities, equalities have been formulated to describe ensemble averages of work and free-energy-related quantities even in far-from-equilibrium situations~\cite{bk,je,cr}. Moreover, it has been demonstrated that in certain systems, it is possible to realize experimentally Maxwell's demon~\cite{dem}, the process in which the information of an observer is used to convert the energy of thermal fluctuations into free energy without performing work on the system~\cite{rmp,mod3}. Qualitatively, such a conversion inverts the information-energy relation in the Landauer principle~\cite{lan} which equates the erasure of information to generation of heat.

Single-electron tunneling (SET) devices which manipulate individual electrons in structures of metallic tunnel junctions \cite{al} have several general advantages for studying nanoscale thermodynamics. These include simplicity of basic dynamics which is understood theoretically to a very good precision,  large degree of experimental control that allows one to design and implement various possible structures, and close connection to practical information processing.  In this paper, we demonstrate that a single-electron pump, monitored by a charge detector which can resolve individual electrons, can be adapted to act as Maxwell's demon, i.e., an information-to-energy converter. We show that in the limit of infinitely fast and error-free detection, the pump can become an ideal energy extractor. This limit can be approached experimentally by lowering the electron tunneling rates in the pump, e.g., by employing hybrid normal-metal--insulator--superconductor (NIS) junctions~\cite{us}.

Away from the ideal limit, the rate and efficiency, with which such an SET demon extracts free energy from thermal fluctuations, depend critically on the detector characteristics: they include the measurement time which is determined by the detector sensitivity and its output noise, and defines the information acquisition rate, and the back-action noise of the detector which should be sufficiently weak to ensure that the demon utilizes only the energy of the thermal fluctuations. The quantitative limitations on the detector obtained below for the demon operation are less stringent than for the quantum-limited detection, the fact that makes realistic the prospect of experimental realization of the Maxwell's demon in the SET configuration.
\begin{figure}
    \includegraphics[width=8.cm]{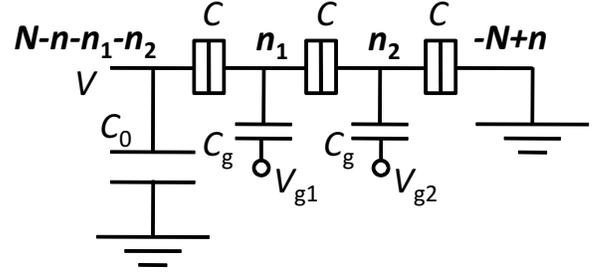}
    \caption{Three-junction electron pump biased by the voltage $V$ that is created by charge $eN$, $N \gg 1$, on a large capacitor $C_0$. The time-dependent gate voltages $V_{g1},V_{g2}$ can move electrons even against the bias from right to left.  The number $n$ counts the electrons transferred through the pump. Electron numbers $n_1,n_2$ on the internal islands of the pump can be monitored by a charge detector (not shown) with sub-electron resolution, e.g., by a single-electron transistor.}
    \label{3j}
\end{figure}

The physical system we consider is the three-junction single-electron pump \cite{pump} shown in Fig.~\ref{3j}. Here, three tunnel junctions define the internal islands of the structure. Each junction has a tunnel resistance $R_T$ and capacitance $C$, and the circuit is controlled by gate voltages $V_{gi}$ on the capacitance $C_g$, at islands $i=1,2$. The pump is biased by the voltage $V$. Although the precise nature of the voltage source is not important, to be specific, we  adopt the model of the source as a large capacitor $C_0\gg C,C_g$ with a large charge $eN$ on it, creating the voltage $V=eN/C_0$ across the pump, where $e$ is the electron charge. We consider the limit $C_0, N \rightarrow \infty$, when the voltage $V$ is independent of the number of electrons $n$ transferred through the pump and the numbers $n_i$ of electrons on the islands $i=1,2$.

The pump operation requires the gate voltages $V_{gi}$ to be time-dependent. This makes it convenient to separate the electrostatic energy $U$ of the system into the gate-voltage-independent part $U_0$ which includes the usual bare charging energy of the pump capacitors and the energy of the charges $n,n_1,n_2$ due to the voltage $V$, and the gate-voltage-dependent bias energy $U_g$, so that $U =U_0 +U_g$. In the assumed regime of small gate capacitances, $C \gg C_g$, we have
\begin{eqnarray}
U_0 = E_C(n_1^2 +n_2^2 +n_1n_2) - eV (n_1+2n_2+3n)/3\, ,
\label{u0} \\
U_g =-E_C[n_1(2n_{g1}+n_{g2})+n_2(n_{g1}+2n_{g2})]\, , \label{A1}
\end{eqnarray}
where $E_C\equiv e^2/3C$ and $n_{gi}\equiv C_gV_{gi}/e$.

\begin{figure}
    \includegraphics[width=8.5cm]{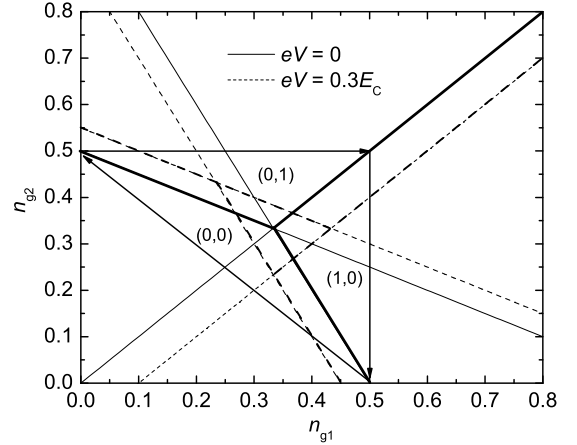}
    \caption{Stability diagram of the pump, and the demon operation cycle. The black solid lines crossing at $n_{g1}=n_{g2}=1/3$ separate the stable charge configurations $(0,0),(1,0)$, and $(0,1)$ of the pump at $V=0$. The dashed lines that form a triangle encircling $n_{g1}=n_{g2}=1/3$ are the corresponding borders for $V=0.3E_C$. Inside this triangle there are no stable charge states, and current runs along the bias. The arrows indicate a possible  trajectory of the demon operation which converts information into free energy.}
    \label{stabd}
\end{figure}

Below, we focus on low-temperature regime, $k_BT \ll E_C$, and gate voltage range such that only three possible charge states on the islands, $(n_1,n_2) = (0,0),(1,0),(0,1)$, can be occupied in practise. The possible changes of the electrostatic energy $U$ in the electron transitions are given then by
\begin{eqnarray}
&& \Delta U_{1,\pm} = \pm E_C(1-2n_{g1}-n_{g2})\mp eV/3\, , \nonumber \\&&
\Delta U_{2,\pm} = \pm E_C(n_{g1}-n_{g2})\mp eV/3\, , \label{A2} \\&&
\Delta U_{3,\pm} = \pm E_C(n_{g1}+2n_{g2}-1)\mp eV/3\, . \nonumber
\end{eqnarray}
Here, the subscript $k,\pm$ refers to tunneling between the three charge states through the $k$th junction from the left end in Fig.~\ref{3j}, with $\pm$ indicating its direction: $+$ to the right, $-$ to the left. The conditions $\Delta U_{k,\pm}=0$ determine the borders of stability of the charge states shown in Fig.~\ref{stabd}:
\begin{eqnarray}
&& (0,0) - (1,0): n_{g2}+2n_{g1} = 1 - eV/3E_C\, ,\nonumber \\&&
(1,0) - (0,1): n_{g2} - n_{g1}=- eV/3E_C\, , \label{A3} \\&&
(0,1) - (0,0): 2n_{g2}+ n_{g1}= 1 + eV/3E_C\, . \nonumber
\end{eqnarray}
For $V=0$, these borders intersect at the triple point $(n_{g1},n_{g2})=(1/3,1/3)$, where all three states have the same energy, or they form a triangle around this point for $V\neq 0$, see Fig.~\ref{stabd}. The standard pumping is obtained~\cite{pump} by changing the gate voltages adiabatically along a trajectory that encircles this triangle.

Thermodynamic properties of the pump follow from the remark that the energy changes $\Delta U_{k,\pm}$ in Eq.~(\ref{A2}) give directly the amount of heat $Q_{k,\pm}$ dissipated in the electrodes in each electron transition \cite{we}, $Q_{k,\pm}=-\Delta U_{k,\pm}$. (Sign change accounts for the natural convention that decreasing energy generates positive heat.) Summing these changes over all electron transitions and adding the change of energy $U$ between the transition due to the time-evolution of $U_g$ (\ref{A1}), we obtain the total change $\Delta U$ over some trajectory of the pump dynamics as
\begin{eqnarray}
\Delta U=-Q -E_C\int \Big[ (2n_1+n_2) dn_{g1}(t) \nonumber \\
+ (n_1+2n_2) dn_{g2}(t) \Big] \, , \label{e11}
\end{eqnarray}
where $Q$ is the total heat generated in the evolution. Subtracting the total change $\Delta U_g$ from both sides of this equation, we obtain the first-law-of-thermodynamics relation for the pump:
\begin{eqnarray}
\Delta U_0=-Q +E_C\int \Big[(2n_{g1}+n_{g2})dn_1(t) \nonumber \\ + (n_{g1}+2n_{g2}) dn_2(t) \Big]\, . \label{e11*}
\end{eqnarray}
Here, the second term on the right coincides with the regular expression for the electrostatic work $W$ done by the changing gate voltages. Similarly, the second term in Eq.~(\ref{e11}) is sometimes used to define the thermodynamic work~\cite{j07} convenient in the discussions of the non-equilibrium fluctuation relations~\cite{je,cr}. Since the charges $n_i$ change by discrete jumps, the work $W$ in Eq.~(\ref{e11*}) reduces to the sum over the times $t_j$ of these jumps
\begin{equation} \label{A5}
W = E_C \sum_j [(2n_{g1}+n_{g2}) \delta n_1+(n_{g1}+2n_{g2}) \delta n_2 ]\Big |_{t=t_j} ,
\end{equation}
where $\delta n_i=\pm 1$ refer to the changes of $n_i$ at the time $t_j$.

We will mostly consider closed cycles of pump evolution, when the gate charges $n_{gi}$ and the island charges $n_i$ return to their initial values at the end of the cycle,
and the only overall change is in the number $n$ of transferred electrons: $\delta n=\pm 1$ for the elementary cycle, depending on the pumping direction, which gives $\Delta U_0 = \pm eV$ from Eq.~(\ref{u0}). Since $k_BT \ll E_C$, we may consider only the cycles that start and end in one certain charge configuration, so that the entropy in the charge degree of freedom of the pump vanishes and free energy $F$ coincides with the internal energy $U_0$ yielding $\Delta F= \Delta U_0$. Using Eq.~(\ref{e11*}) we get for pumping against the bias
\begin{equation} \label{e12}
W = \Delta U_0+Q=eV+Q \, ,
\end{equation}
i.e., the work done by the gates in one cycle is partly dissipated into heat and partly increases the free energy of the pump by charging the capacitor $C_0$.

As an application of Eq.~(\ref{e12}), we first consider the usual slow pumping, for which there is no need to extract information on the system. Because of the randomness of the tunneling events, the heat $Q$, and therefore the work done by the gates, fluctuate from cycle to cycle. Statistics of $Q$ and $W$ over the cycles can be determined from the statistics of dissipated heat in the individual SET transitions from one charge state to the next in the cycle. Quantitatively, this statistics depends on the electron tunneling rates $\Gamma_{k,\pm}$ for the transitions between the charge states in Eq.~(\ref{A2}), which, in the case of the pump made of normal-metal conductors with constant density of states, are given by the standard expression $\Gamma =(e^2R_T)^{-1}\Delta U/[e^{\Delta U /(k_BT)}-1]$. In the limit of slow evolution, the distribution of heat in the individual SET transitions in this case is known to be Gaussian \cite{we}. The sum of the Gaussian distributions in the three transitions of one cycle  results in the Gaussian distribution of the total heat $Q$, with the average $\langle Q\rangle$ and the width $\sigma_Q$ of the distribution related in the same way as for the individual transitions,
$\sigma_Q^2 = 2k_B T \langle Q\rangle$~\cite{we}.
The average $\langle Q\rangle$ depends on the rates $\eta_k$ of the change of energies $\Delta U_{k,\pm}$ in Eq.~(\ref{A2}) at the time when the gate voltage trajectory crosses the border between the charge states corresponding to the transition in the $k$th junction (Fig.~\ref{stabd}): $\langle Q\rangle \simeq 0.43 (e^2 R_T /k_BT) \sum |\eta_k|$. For instance, in a typical case of harmonic variation of gate voltages with frequency $\omega$ and amplitude $V_g$, we have $n_{g1}(t) = 1/3+(C_gV_g/e)\cos (\omega t)$, $n_{g2}(t) = 1/3-(C_gV_g/e)\sin (\omega t)$, and for $eV\ll E_C$,
\[ \langle Q\rangle \simeq 2.51 (e V_g E_C /k_BT) C_gR_T \omega\, . \]
Thus, for $\omega \rightarrow 0$, the generated heat vanishes in every pumping cycle, and $W=eV$, i.e., the work done by the gates on the tunneling electrons to pump them against the bias equals the change in free energy.

Let us consider next the demon-type operation, in which electrons flow against the bias with no net work done by the gate voltages, $W=0$. As follows from Eq.~(\ref{e12}), in this regime, $\Delta F=eV = -Q$,  i.e., $Q<0$,  and the energy is extracted from thermal fluctuations. This is achieved by a detector that can observe transitions in the pump. A possible operation cycle of the demon is described on the stability diagram in Fig.~\ref{stabd} by the triangle of arrows. First, set $n_{g1}=0, n_{g2}=1/2$, i.e., the degeneracy point at $V=0$ for the charge states $(0,0)$ and $(0,1)$, and take $(0,0)$ to be the initial state of the  cycle. Comparison of the two energy differences in Eq.~(\ref{A2}) at these gate voltages, $\Delta U_{1,+}=E_C/2 -eV/3$ and $\Delta U_{3,-}=eV/3$, shows that for $E_C \gg 4eV/3 \sim k_BT$ the system will make with dominant probability a fluctuation-induced transition into the state $(0,1)$ (against the bias) and not to $(1,0)$ (along the bias). As soon as the transition to $(0,1)$ occurs, the detector registers it and the gate voltages shift quickly to the position $n_{g1}=n_{g2}=1/2$. Similar arguments as above show that at these values of the gate voltages, with dominant probability, the system makes the transition to state $(1,0)$. Once the transition is observed, the gates are moved quickly to $n_{g1}=1/2, n_{g2}=0$,  and when the next dominant transition to $(0,0)$ occurs and is registered, the gates move back to the initial positions $n_{g1}=0, n_{g2}=1/2$, completing the cycle. Equation (\ref{A5}) can be used to show that the total work done in this process vanishes, $W=0$. (The work in the first and last transitions sum up to zero and vanishes for the middle transition.) However, the tunneling electron charges the capacitor $C_0$ increasing the free energy of the pump by $\Delta F =eV$ in the closed cycle with the energy coming from thermal fluctuations, $Q=-eV<0$. The average generated power by the demon is $P=(eV/3) \Gamma (eV/3)$.

The central element of the demon operation described above is the measurement and feedback control sequence which needs to be much faster than the electron tunneling rates. The typical charge detectors used in mesoscopic structures like the SET pump here are based on either the SET transistors~\cite{al} or quantum point contacts~\cite{qpc}. Both types of detectors convert the input charge signal $q$ of a sub-electron magnitude into the output signal of the dc current $I$, effectively amplifying the input. For weak detector coupling to the pump, the conversion process is linear and can be characterized by the linear-response coefficient $\lambda=\partial I/\partial q$. General statistical mechanics of the detectors shows that the measurement is unavoidably accompanied by the output current noise with some spectral density $S_I$ and the back-action noise $S_V$ of the potential of the island coupled to the  detector, since the spectral densities are limited from below by what is effectively the Heisenberg uncertainty relation, $S_I S_V \geq (\hbar \lambda/4\pi)^2$ (see, e.g., \cite{det}). Both spectral densities are constant below a cut-off frequency $\omega_C$ of the classical part of the noise. We focus the subsequent discussion on the case of SET electrometers, although qualitatively, the same considerations apply to measurements by quantum point contacts. The charges on the two islands should be either monitored by two independent detectors or by a single detector coupled asymmetrically to both
islands. Here, the individual detector parameters can be different, but the equations below still apply to the
detection process of each individual transition.

The main effect of the back-action noise $S_V$ of the potential is the transfer of energy to electrons tunneling in the pump, stimulating the tunneling rate. Quantitatively, this effect can be described as decoherence of the voltage drop across the tunnel junction. The white noise $S_V$ produces the simplest exponential decoherence with the rate $\gamma' =\pi e^2 S_V/\hbar^2$. In the usual description of the rate $\Gamma$ of electron tunneling between two normal-metal electrodes, this decoherence gives the Lorentzian lineshape of the tunneling,
 \[\Gamma= \frac{1}{e^2R_T}\int d \epsilon_i d\epsilon_f f (\epsilon_i) [1-f (\epsilon_f)] \frac{\gamma/\pi}{(\epsilon_i -\epsilon_f-\Delta U)^2 +\gamma^2}\, , \]
where $f(\epsilon)$ is the Fermi distribution of electrons in the electrodes, and $\gamma\equiv \hbar \gamma'$. In operating Maxwell's demon, one would be interested in extracting energy from thermal fluctuations, and not from the detector noise. Thus it is important to keep the effects of noise small. For small but finite $\gamma$, evaluating the tunneling rate, one obtains
$\Gamma= \frac{1}{e^2R_T} [\frac{\Delta U}{\exp[\Delta U/(k_BT)]-1}+ \frac{\gamma}{\pi} \ln (\hbar \omega_C/k_BT) ]$ ,
where the second term represents the detector-induced correction with logarithmic accuracy in large frequency $\omega_C$. For the SET detector, this frequency is roughly given by the bias voltage $V_D$ across the transistor, $\omega_C\simeq eV_D/\hbar$. The condition that this correction is small for $\Delta U\sim k_BT$ (regime of demon operation) imposes constraint on $S_V$:
\begin{equation} \label{e7}
S_V \ll \hbar k_BT/[e^2 \ln (\hbar \omega_C/k_BT) ] \, .
\end{equation}

The cut-off frequency $1/\tau$ of the output detector noise $S_I$ is typically much smaller than the intrinsic cut-off $\omega_C$ because of the finite bandwidth of the information carrying wiring.
For instance, a standard dc coupled SET electrometer has a bandwidth of about 1 kHz~\cite{us}. (In principle, it can be improved to the megahertz range with an rf-SET electrometer~\cite{rf}.) Together with $S_I$, the  cut-off $1/\tau$ determines the accuracy of detection of the measured charge state. Assuming the usual Lorentzian lineshape of the spectrum, $S_I(\omega)=S_I(0)/[1+(\omega \tau)^2]$, we estimate the magnitude of the output current noise as $\sigma \equiv \langle \tilde{I}^2\rangle =\pi S_I(0)/\tau$. Distinguishing the two values of the output current that correspond to the two charge states, electron on or off the island,
that differ by $\Delta I=e\lambda$, in the presence of this noise can be achieved only with a finite probability $p$ of mistake. The output current noise is produced by many tunneling electrons in the transistor and is therefore Gaussian. This implies that
$ p =(1/2) \mbox{erfc}(\Delta I/\sqrt{8\sigma}) \simeq (\sqrt{2\sigma/\pi} /\Delta I) e^{-(\Delta I)^2/8\sigma}$.
The condition that mistakes in the charge detection are negligible, imposes the following constraint:
\begin{equation} \label{e8}
S_I/\lambda^2 \ll e^2\tau/8\pi \, .
\end{equation}
Since the detector makes the choice of two options, electron on or off the island, the amount of usable information obtained in measurement of each transition is roughly given by $\ln 2$, while the energy extracted is $eV/3\simeq k_B T$. We see that the SET demon has qualitatively the same information/energy relation as in the standard Landauer principle \cite{lan}, with one bit of information corresponding roughly to $k_B T$ of free energy gain.

If the condition (\ref{e8}) is satisfied, the time required to detect the change in the charge state of the pump is given directly by $\tau$. The operation of the Maxwell's demon described above assumed that electron transition is detected, and the gate voltages are changed before the electron has a chance to tunnel back, i.e.,
$\Gamma \ll \tau^{-1}$, where $\Gamma$ is the rate of such reverse tunneling. This tunneling is driven by the energy change $eV/3$, so that $\Gamma \sim V/(eR_T)$ for the normal-metal pump. The conditions (\ref{e7}), (\ref{e8}), and $\Gamma \ll \tau^{-1}$ are compatible with the Heisenberg relation for detectors, $S_I S_V/\lambda^2 \geq (\hbar /4\pi)^2$, provided that $\hbar \Gamma \ll k_BT$, and thus they can be satisfied with the detector far from being quantum-limited (which satisfies Heisenberg relation as equality). Practical limitations on the detector bandwidth, however, imply that these conditions cannot be satisfied with usual normal-metal pumps. Indeed, typically $E_C/e \gg$ 100 $\mu$V. In order to exploit this available energy one would like to set $V \sim 100 $ $\mu$V, which would be reasonably larger than $k_BT$ at the typical working temperature of about 100 mK for such a device. Even if the detector is fast, $\tau \sim 1$ $\mu$s, to satisfy condition $\Gamma \ll \tau^{-1}$ we need to have $R_T \gg V\tau/e \sim 1$ G$\Omega$, which is impractically large.
This problem can be resolved by employing superconducting and normal-metal electrodes at each junction of the pump, i.e., creating NIS junctions~\cite{us}. In this case~\cite{ns1}, the rates at $eV/3 \ll \Delta$, where $\Delta$ is the superconducting gap, are suppressed exponentially from those of the normal state junctions, and it is easy to satisfy $\Gamma \ll \tau^{-1}$ with $R_T\sim 1$ M$\Omega$, e.g., at the temperature of 100 mK~\cite{saira10}. Condition on the negligible back-action is also satisfied more easily in this case by making $\hbar \omega_C \ll \Delta$, so that the back-action noise does not excite tunneling. Simple estimates of $\lambda$ and $S_I$ for the SET transistor show that Eq.~(\ref{e8}) can also be satisfied by taking the tunneling rates $\Gamma_D$ in it and its coupling capacitance $C_0$ to the pump such that $\Gamma_D \tau (C_0/C)^2\gg 8\pi$. With the discussed pump parameters, the power generated by the demon is limited to $P \lesssim 10$ aW, and is hardly observable directly, but only indirectly by measuring the collected charge at the bias capacitor $C_0$.

In conclusion, we have designed a scheme to realize Maxwell's demon based on a single-electron pump and charge detectors. We derived conditions that the pump and detectors should satisfy for demon operation. Only the pumps based on the NIS junctions, i.e., with superconducting and normal-metal electrodes at each junction,
seem to provide an experimentally feasible device to satisfy these conditions.

\end{document}